%% file: main.tex
\PassOptionsToPackage{unicode}{hyperref}
\PassOptionsToPackage{hyphens}{url}
\PassOptionsToPackage{dvipsnames,svgnames,x11names}{xcolor}
\documentclass[
  12pt]{article}

\usepackage{amsmath,amssymb}
\usepackage{iftex}
\ifPDFTeX
  \usepackage[T1]{fontenc}
  \usepackage[utf8]{inputenc}
  \usepackage{textcomp} 
\else 
  \usepackage{unicode-math}
  \defaultfontfeatures{Scale=MatchLowercase}
  \defaultfontfeatures[\rmfamily]{Ligatures=TeX,Scale=1}
\fi
\usepackage{lmodern}
\ifPDFTeX\else  
\fi
\IfFileExists{upquote.sty}{\usepackage{upquote}}{}
\IfFileExists{microtype.sty}{
  \usepackage[]{microtype}
  \UseMicrotypeSet[protrusion]{basicmath} 
}{}
\makeatletter
\@ifundefined{KOMAClassName}{
  \IfFileExists{parskip.sty}{%
    \usepackage{parskip}
  }{
    \setlength{\parindent}{0pt}
    \setlength{\parskip}{6pt plus 2pt minus 1pt}}
}{
  \KOMAoptions{parskip=half}}
\makeatother
\usepackage{xcolor}
\makeatletter
\ifx\paragraph\undefined\else
  \let\oldparagraph\paragraph
  \renewcommand{\paragraph}{
    \@ifstar
      \xxxParagraphStar
      \xxxParagraphNoStar
  }
  \newcommand{\xxxParagraphStar}[1]{\oldparagraph*{#1}\mbox{}}
  \newcommand{\xxxParagraphNoStar}[1]{\oldparagraph{#1}\mbox{}}
\fi
\ifx\subparagraph\undefined\else
  \let\oldsubparagraph\subparagraph
  \renewcommand{\subparagraph}{
    \@ifstar
      \xxxSubParagraphStar
      \xxxSubParagraphNoStar
  }
  \newcommand{\xxxSubParagraphStar}[1]{\oldsubparagraph*{#1}\mbox{}}
  \newcommand{\xxxSubParagraphNoStar}[1]{\oldsubparagraph{#1}\mbox{}}
\fi
\makeatother

\usepackage{longtable,booktabs,array}
\usepackage{calc} 
\usepackage{etoolbox}
\makeatletter
\patchcmd\longtable{\par}{\if@noskipsec\mbox{}\fi\par}{}{}
\makeatother
\IfFileExists{footnotehyper.sty}{\usepackage{footnotehyper}}{\usepackage{footnote}}
\makesavenoteenv{longtable}
\usepackage{graphicx}
\makeatletter
\def\maxwidth{\ifdim\Gin@nat@width>\linewidth\linewidth\else\Gin@nat@width\fi}
\def\maxheight{\ifdim\Gin@nat@height>\textheight\textheight\else\Gin@nat@height\fi}
\makeatother
\setkeys{Gin}{width=\maxwidth,height=\maxheight,keepaspectratio}
\makeatletter
\def\fps@figure{htbp}
\makeatother

\makeatletter
\@ifpackageloaded{caption}{}{\usepackage{caption}}
\AtBeginDocument{%
\ifdefined\contentsname
  \renewcommand*\contentsname{Table of contents}
\else
  \newcommand\contentsname{Table of contents}
\fi
\ifdefined\listfigurename
  \renewcommand*\listfigurename{List of Figures}
\else
  \newcommand\listfigurename{List of Figures}
\fi
\ifdefined\listtablename
  \renewcommand*\listtablename{List of Tables}
\else
  \newcommand\listtablename{List of Tables}
\fi
\ifdefined\figurename
  \renewcommand*\figurename{Figure}
\else
  \newcommand\figurename{Figure}
\fi
\ifdefined\tablename
  \renewcommand*\tablename{Table}
\else
  \newcommand\tablename{Table}
\fi
}
\@ifpackageloaded{float}{}{\usepackage{float}}
\floatstyle{ruled}
\@ifundefined{c@chapter}{\newfloat{codelisting}{h}{lop}}{\newfloat{codelisting}{h}{lop}[chapter]}
\floatname{codelisting}{Listing}

\makeatother
\makeatletter
\@ifpackageloaded{caption}{}{\usepackage{caption}}
\@ifpackageloaded{subcaption}{}{\usepackage{subcaption}}
\makeatother

\ifLuaTeX
  \usepackage{selnolig}  
\fi
\usepackage[]{natbib}
\usepackage{bookmark}

\IfFileExists{xurl.sty}{\usepackage{xurl}}{} 
\hypersetup{
  pdftitle={Predicting Bot Vulnerability from Posting Trajectories: Censored Functional Regression under Informative Sampling},
  pdfauthor={Jake Koerner; Ana-Maria Staicu; Caitrin Murphy; Eric Laber; Mihai Nicola; William Rand; Zakaria Babutsidze},
  pdfkeywords={Censored Functional Data, Functional Prediction Models, Informative Time Designs, Matched case-control, Social media data},
  colorlinks=true,
  linkcolor={blue},
  filecolor={Maroon},
  citecolor={Blue},
  urlcolor={Blue},
  pdfcreator={LaTeX via pandoc}}

\usepackage[table]{xcolor} 
\usepackage[export]{adjustbox}
\usepackage{longtable}    
\newcommand{\anon}{1}

\begin{document}
\include{commands}

\def\spacingset#1{\renewcommand{\baselinestretch}%
{#1}\small\normalsize} \spacingset{1}


\if1\anon
{
 \title{\bf Predicting Bot Vulnerability from Posting Trajectories: Censored Functional Regression under Informative Sampling}

\author{
Jake Koerner
\hspace{.2cm}\\
North Carolina State University\\
\and
Ana-Maria Staicu\\
North Carolina State University\\
\and
Caitrin Murphy\\
Duke University\\
\and
Eric Laber\\
Duke University\\
\and
Mihai Nicola\\
Stevens Institute of Technology\\
\and
William Rand\\
North Carolina State University
\and
Zakaria Babutsidze\\
SKEMA Business School
}

\maketitle
} \fi

\if0\anon
{
  \bigskip
  \bigskip
  \bigskip
  \begin{center}
    {\LARGE\bf Predicting Bot Vulnerability from Posting Trajectories: Censored Functional Regression under Informative Sampling}
\end{center}
  \medskip
} \fi

\bigskip
\begin{abstract}
In this manuscript, we propose a novel framework for Scalar-on Censored Informative-design Functional Regression or SoCIFR. This setting is increasingly common in modern longitudinal and digital data applications but remains underdeveloped in functional data literature. We first discuss estimation and prediction in SoCIFR and extend the methodology to accommodate a matched case-control design. The proposed methodology is further generalized to handle multiple functional predictors, allowing for both censored and uncensored trajectories, observed under informative or non-informative sampling designs. Through simulation studies, we assess the performance of the proposed methods under various data-generating scenarios. We apply the methods to the motivating application for predicting user susceptibility to automated ("bot") interactions at increasing future time horizons based on social media behavioral trajectories observed over fixed time windows.
\end{abstract}

\noindent%
\vfill

\newpage
\spacingset{1.8} 

%

\section{Introduction}\label{sec-intro}

Online social media platforms such as X/Twitter and Facebook have transformed global communication, but they have also increased users' exposure to automated accounts that disseminate misleading or manipulative content. These bot accounts are often difficult to identify (\cite{ferrara2016rise}) and have been implicated in the spread of political disinformation (\cite{Ferrara_2017,Badawy_2018}), false medical claims (\cite{jamison2019malicious}), and COVID-19 misinformation (\cite{owen2020}).

Considerable research has focused on detecting automated accounts using a variety of methodologies (see \cite{RODRIGUEZ_2020} for a review). More recently, attention has shifted toward understanding human--bot interactions and their effects on user behavior. Prior studies show that interactions with bots can amplify extreme views, influence sentiment, and alter online engagement, particularly in political and activist settings. For example, \cite{Yan2023} demonstrate that users who identify bot activity tend to overestimate both the prevalence of bots and their influence on others. \cite{Li2024SocialBS} show that interaction with bots negatively affects sentiment toward activism while maintaining engagement levels, whereas \cite{Askari2024} find that bot interactions primarily increase engagement among users who are already highly politically active. \cite{Li2026} report that repeated exposure to content labeled as AI-generated reduces users’ emotional responses over time.

Relatively little work has focused on predicting a user's susceptibility to interacting with bot accounts. Previous studies have examined behavioral correlates of susceptibility (\cite{wagner2012}) and the influence of bots within users' social networks (\cite{Aldayel2022}), but neither addressed how longitudinal user behavior predicts future bot interactions. \cite{Babutsitdze2022} found that users exhibit changes in engagement and posting behavior around the time of bot interaction, including shifts in posting frequency, post length, and mentions. However, that study did not directly model future susceptibility or compare longitudinal trajectories between susceptible and non-susceptible users. The present work fills this gap by leveraging repeated posting behavior over time to predict future interactions with bot accounts.

Our work is motivated by a matched case-control Twitter/X dataset consisting of users who interacted with bots and matched control users who did not. User activity is summarized longitudinally through posting features such as posting volume, post length, and mention (@) behavior over the year preceding bot interaction (or the corresponding period for controls). To capture stable behavioral patterns, daily activity is aggregated to the weekly level and represented using functional data analysis techniques. The resulting longitudinal features include both censored and informatively missing components arising from periods of user inactivity and incomplete mention information. To account for these complexities, we introduce a latent activity process and develop interpretable models for predicting susceptibility to future bot interaction from repeated posting behavior over time.

Scalar-on-function models with generalized responses are well established in the functional data analysis literature \citep{James:2002, muller2005generalized, goldsmith2011penalized, McLeanEtAl2012, GertheissMaityStaicu2013, UssetStaicuMaity2016}. In recent years, attention has turned to censored functional data \citep{murphy2026, dey2024functional} and informative observation designs \citep{Weaver2023, Xu2022, sang2024}, though these challenges have largely been studied separately. In our motivating application, both arise simultaneously, and ignoring either can lead to biased inference. We develop functional regression models that jointly accommodate censored longitudinal signals and informative observation processes through latent process modeling and inverse-probability weighting.

We extend the censored functional data framework of \cite{murphy2026} by introducing subject-specific latent processes that jointly characterize the censored trajectories and the informative observation mechanism governing posting activity. Unlike existing approaches, our framework explicitly accommodates missing-not-at-random (MNAR) sampling while jointly estimating both latent processes. The proposed approach shares similarities with sliced inverse regression \citep{li1991sir} in that it extracts information about the informatively observed, potentially censored predictor process through the conditional distribution of the response. We then use a multivariate functional data analysis approach to extract low-dimensional features that capture the dominant modes of variation in both the observed censored signals and the latent activity process. Additional functional covariates, when available, are incorporated within the same framework to obtain unified subject-specific representations. Building on these representations, we develop scalar-on-function regression models for binary outcomes with censored and informatively sampled functional predictors.  We further extend the methodology to matched case-control settings \citep{breslow1980statistical}, motivated by our application on predicting susceptibility to bot interaction. Accounting for the matched design is important when the matching variables are associated with the outcome, as ignoring the matching structure may lead to biased estimation and prediction.

Our work makes several key contributions. (1) We introduce a unified modeling framework for censored functional data under structural missingness and MNAR mechanisms that explicitly incorporates the missingness process into estimation. (2) We develop scalar-on-function regression methods for binary outcomes with censored functional predictors observed under informative sampling designs. (3) We extend the proposed framework to matched case-control settings. (4) In our motivating application, the methodology enables prediction of a user’s future susceptibility to bot interaction over a specified prediction horizon using repeatedly observed posting features collected over a recent observation window. (5) We provide open-source software implementing the proposed methods, facilitating reproducibility and broader application to censored or irregularly sampled functional data settings (https://github.com/jrkoerne/botSusceptibility).

The remainder of the paper is organized as follows. Section~\ref{sec:data_application} introduces the motivating application. Section~\ref{sec:framework} presents the statistical framework and proposed methodology. Section~\ref{sec:extensions} extends the approach to multiple functional predictors under both informative and non-informative observation schemes, and to matched case-control designs. Section~\ref{sec:simulations} evaluates finite-sample performance through simulations and Section \ref{sec:dataApp} applies the proposed methods to the motivating dataset. Section~\ref{sec:discussion} concludes with a discussion.

\section{Data Application Description} \label{sec:data_application}

The motivating application is a matched-pair study of Twitter/X users, where bot-interacting users were matched to controls on account metadata (US state and creation year). For each user, daily posting features, including post volume and average length, and total @{ } mentions, were recorded over the year preceding the first bot interaction (or the corresponding period for controls). See \cite{Babutsitdze2022} for a full study description.

The dataset consists of 12,180 matched pairs of Twitter/X users from across the United States. Users were classified as bot-interacting if they liked, replied to, or mentioned one of 51 influential bot accounts identified among the 2,752 malicious accounts removed by Twitter/X between 2011 and 2017. Daily posting features were aggregated to the weekly level. Weekly post volume was defined as $\log(x+1)$, where $x$ is the average daily number of posts, and weekly tweet length as the average daily ratio of words to tweets. Because post-level mention indicators were unavailable, the ``@''-tweet rate was approximated by the daily ratio of total mentions to total posts, capped at $1$, and averaged within each week. Tweet length and ``@''-tweet rate were observed only during active weeks and treated as missing otherwise, while post volume was recorded as $0$.

After excluding users associated with two pre-2015 bot accounts, users from sparsely represented states, and pairs with degenerate posting behavior, 9,586 matched pairs remained for analysis. Our goal is to predict susceptibility to bot interaction from longitudinal trajectories of tweet volume, capped ``@''-tweet rate, and average tweet length while accounting for informative activity patterns and the matched-pair design.

\begin{figure}[H]
    \centering
    \includegraphics[width=\linewidth]{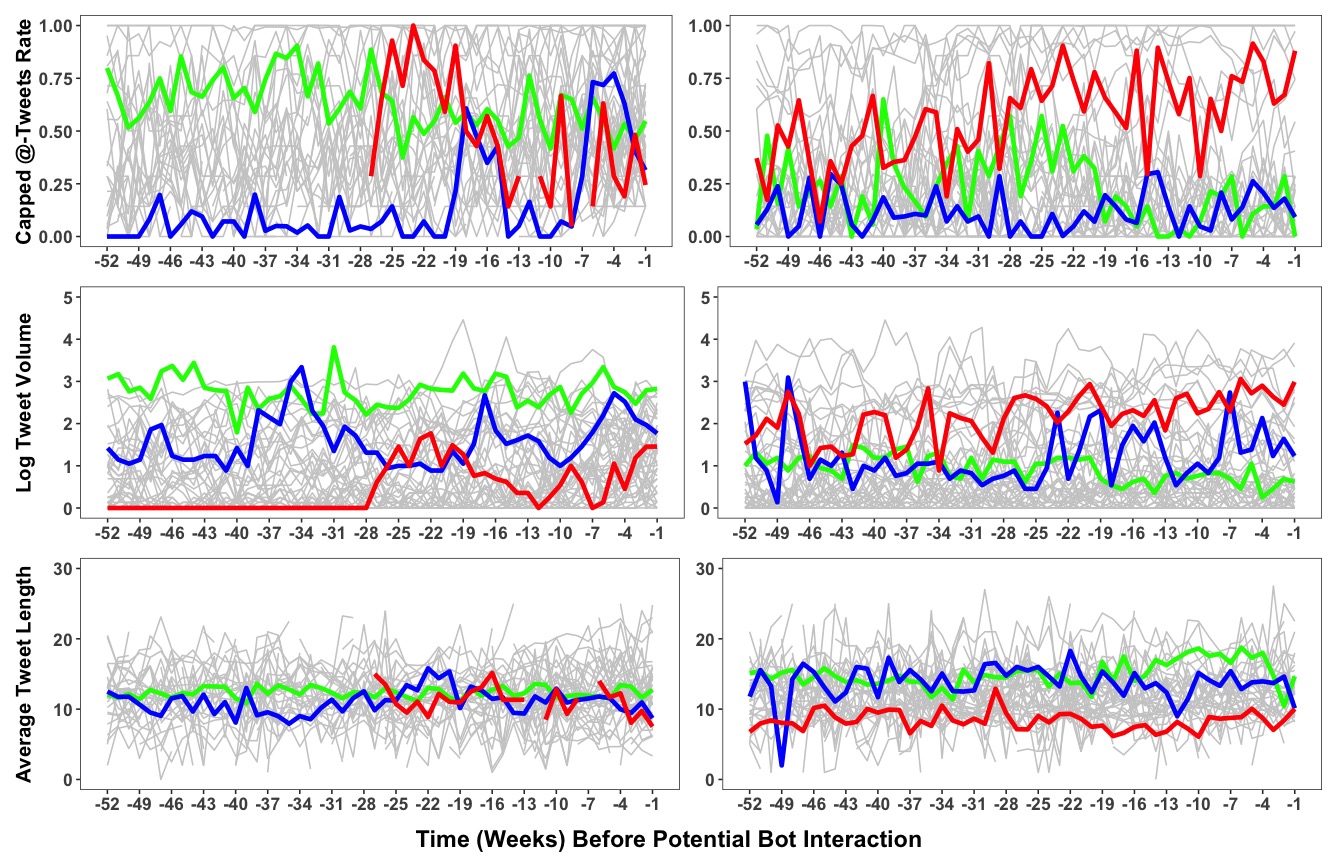}
    \caption{Weekly ``trajectories" of the proportion of tweets containing “@” (top), log tweet volume (middle), and average tweet length (bottom) over a 52-week period for 50 susceptible users (left) and 50 matched controls (right). Trajectories for three randomly selected matched pairs are highlighted. Consecutive weekly observations for each user are connected.
    }
    \label{fig:allData}
\end{figure}

Figure~\ref{fig:allData} shows trajectories of capped ``@''-tweet rate, log tweet volume, and average tweet length for susceptible users and matched controls during the year preceding bot interaction. The figure suggests substantial heterogeneity in posting behavior as well as systematic differences between the two groups, particularly in ``@''-mention activity. Supplementary Figure~S.1 further shows increasing posting activity among susceptible users as the interaction week approaches, whereas controls remain relatively stable over time.

Preliminary analyses indicate negligible correlation in posting behavior within matched pairs (see Supplementary Material). Accordingly, we develop the methodology under a working independence assumption and propose a framework for extracting features from censored longitudinal signals observed under informative sampling. These features are then used to construct interpretable models for predicting susceptibility to bot interaction, with a subsequent extension to matched case-control studies.

\section{Statistical Framework and Estimation Methodology}\label{sec:framework}

\subsection{Scalar-on Censored Informative-Design Functional Regression} \label{ssec_Meth1}

Suppose user $i$ contributes data $\{(W_{ij},t_{ij})_{j=1}^{m_i},Y_i\}$, where $W_{ij}\in[a,b]$ is a bounded repeated measurement observed at informative times $t_{ij}\in\mathcal{T}$ and $Y_i\in\{0,1\}$ is a binary outcome. We refer to $\{(W_{ij},t_{ij})\}_{j=1}^{m_i}$ as \emph{censored functional data observed under an informative sampling design}. The goal is to extract informative low-dimensional features for prediction. Existing methods address either censored functional predictors \citep{murphy2026} or informative observation processes \citep{Gervini2020, Weaver2023, sang2024}, but not both simultaneously. We therefore develop a two-part framework that jointly models the censoring mechanism and the observation process.

\emph{Part I (Censoring mechanism)}. To capture temporal dependence, we introduce a latent stochastic process $Z_i(\cdot)$ on $\mathcal{T}$ with smooth mean and covariance functions and assume the observed measurements $W_{ij}$ satisfy
\begin{equation}
W_{ij}
=
\begin{cases}
a, & Z_i(t_{ij}) + \epsilon_{ij} < a, \\
Z_i(t_{ij}) + \epsilon_{ij}, & Z_i(t_{ij}) + \epsilon_{ij} \in [a,b], \qquad j=1,\ldots,m_i, \\
b, & Z_i(t_{ij}) + \epsilon_{ij} > b,
\end{cases}
\label{eqn_modelW}
\end{equation}
where $\epsilon_{ij}$ denotes measurement error. The errors $\epsilon_{ij}$ are assumed to be independent and identically distributed with mean zero across both subjects $i$ and measurement occasions $j$. 

\emph{Part II (Informative observation process)}. Let $R_i(\cdot)$ denote a binary observation process on $\mathcal{T}$, where $R_i(t_\ell)=1$ if user $i$ is observed at time $t_\ell$ and $0$ otherwise. Following \citet{hall2008modelling}, we model temporal dependence in $R_i(\cdot)$ through a latent process $X_i(\cdot)$ and assume that, conditional on $X_i(\cdot)$,
\begin{equation}
R_i(t_\ell) \mid X_i(\cdot), t_\ell 
\;\overset{\text{ind}}{\sim}\;
\mathrm{Bernoulli}\{p_i(t_\ell)\}, 
\qquad 
g\{p_i(t_\ell)\}=X_i(t_\ell),
\label{eqn:modeling_timepoints_1}
\end{equation}
for $t_\ell \in \mathcal{T}$, where $g(\cdot)$ is a known monotone differentiable link function, $X_i$s are assumed iid with smooth mean and covariance functions, and satisfy $g^{-1}\{X_i(t)\}>0$ for all $t\in\mathcal{T}$.

Furthermore, we assume that $\{\epsilon_{ij}: j=1,\ldots,m_i\}$ are independent of the observation process $R_i(\cdot)$ and the latent observation-time process $X_i$; i.e. $(W_{i1}, \ldots, W_{im_i}) \perp (R_i,X_i) \,\big|\, Z_i$ or assumption (A1). In addition, conditional on the latent process governing the time observation mechanism, we assume that the time observation process is independent of the latent process underlying the censoring mechanism; that is, $ R_{i} \perp Z_i  \,\big|\,  X_i $ (A2).

We assume that, for the purpose of predicting the class outcome $Y_i$, all relevant information in the observed data is captured by the latent processes $X_i(\cdot)$ and $Z_i(\cdot)$. Formally, we assume the conditional independence $
Y_i \;\perp\; (R_i, W_{i1}, \ldots, W_{im_i}) \,\big|\, (Z_i,X_i)$ or assumption (A3).
Under this assumption, the response $Y_i$ is modeled as a function of the latent processes $Z_i(\cdot)$ and $X_i(\cdot)$ through
$
h\{ \Pr(Y_i = 1 \mid Z_i, X_i) \} = f(Z_i, X_i),
$
where $h(\cdot)$ is a known link function and $f(\cdot,\cdot)$ is an unknown functional characterizing the joint effect of the latent processes, allowing for linear, additive, or more general non-additive structures. It follows that if $\bzeta_i$ is a lower dimensional vector of features of the pair processes $Z_i(\cdot)$ and $X_i(\cdot)$, then  $h \left \{P(Y_i=1|Z_i, X_i) \right \} \approx \beta_0 + \boldsymbol{\beta}^T \bzeta_i$ for unknown $\beta_0$ and $\boldsymbol{\beta}$. Other generalized functional regression can be considered using the vector of lower dimensional features $\bzeta_i$; we include such cases in simulations and the data application. 

Our estimation approach relies on the following observation. Denote $p^c_{i}(t)={E[R_i(t)|X_i, Y_i=c]}$; for any $t\in\mathcal{T}$ we have that $E\left [Z_i(t)R_i(t) \frac{1}{p^c_{i}(t)}\ \Big|  Y_i=c \right ]$ equals
\beqn
&& E_{X_i|Y_i=c}\left \{E \left [Z_i(t) R_i(t)\frac{1}{p^c_{i}(t)} \ \Big|  \ X_i,Y_i=c \right ]  \right \}  \nonumber \\ 
&&= E_{X_i|Y_i=c} \left \{\frac{1}{p^c_{i}(t)} E\left [Z_i(t) | X_i, Y_i=c \right ]
E\left [R_i(t) | X_i, Y_i=c \right ]  \right \}= E[Z_i(t) \ | Y_i=c]. \label{eqn:mean_identifiability}
\eeqn
The overall roadmap for the estimation methodology has three main components:

\begin{itemize}
    \item[1.] Estimate the class-dependent observation process at the user level, $p_i^c(t)$;
    \item[2.] Recover the latent process $Z_i(\cdot)$ from censored data under informative sampling;
    \item[3.] Extract low-dimensional features for prediction.
\end{itemize}

\subsection{Censored Functional Regression
under Informative Sampling} \label{sec:estimation} 

\subsubsection{Recovering the latent observation process}\label{ssecE1}

Recall that $\{t_1,\ldots,t_m\}$ denotes the set of observation times across all subjects, and let $R_{il}$ indicate whether subject $i$ is observed at time $t_l$. Under model (\ref{eqn:modeling_timepoints_1}), conditional on $X_i(\cdot)$, the $R_{il}$ are independent Bernoulli variables with
$
P(R_{il}=1)=g^{-1}\{X_i(t_l)\}.
$
We estimate the user-specific latent processes separately within each class to accommodate class-dependent observation patterns, and subsequently estimate their marginal mean and covariance functions. Although related in spirit to sliced inverse regression \citep{li1991sir}, our feature extraction approach uses the full data rather than sliced subsets (see Section \ref{ssecE3}).

Let $c\in\{0,1\}$ index account type and denote by $X_i^c(\cdot)$ the latent observation process $X_i(\cdot)\mid Y_i=c$. Following \citet{hall2008modelling}, we assume that $X_i^c(\cdot)$ deviates only modestly from its class-specific mean, allowing a first-order Taylor expansion of $g^{-1}\{X_i(t)\}$ around the mean. This approximation links the mean and covariance of the latent process to those of the observed binary process. Define
$
\alpha_c(t)=E\{R_i(t)\mid Y_i=c\}$ and $\beta_c(t,t')=E\{R_i(t)R_i(t')\mid Y_i=c\}.
$
These quantities can be estimated by smoothing the binary observations $\{R_{il}:Y_i=c\}$ and their pairwise products $\{R_{il}R_{il'}:Y_i=c\}$ over time. The class-specific mean $\mu_X^c(\cdot)$ and covariance $\Sigma_X^c(\cdot,\cdot)$ of $X_i^c(\cdot)$ are then estimated by
\beqn 
\widehat \mu^c_{X}(t) &= &g \left \{ \widehat {\alpha}_c(t) \right \} \\
\widehat {\Sigma}^c_{X}(t, t') &= &\frac{ \widehat {\beta}_c (t,t') - \widehat {\alpha}_c(t)\widehat {\alpha}_c(t') }{ (g^{-1})^{'}\{\widehat \mu^c_{X}(t)  \} (g^{-1})^{'}\{\widehat \mu^c_{X}(t')  \}  }.
\label{eqn:estimatingmean_cov_X_group}
\eeqn
Next, we represent $X_i^c(\cdot)$ using a truncated \KL (KL) expansion and predict the corresponding scores via a generalized linear mixed model \citep{gertheiss2017note}. Let $\{\widehat \lambda^c_{X,k}, \widehat \psi^c_{X,k}(\cdot) \}_k$ denote the eigenvalue - eigenfunction pairs of $\widehat{\Sigma}^c_X(\cdot ,\cdot)$, and let $K_{X,c}$ be a truncation level. The latent scores are estimated using the model
\beqn
R_{il} \Big | Y_i=c  &\overset{ind}{\sim} &Bernoulli (p^c_{i}(t_l)), \ \text{for} \ l=1, \ldots, m \nonumber \\
g \{ p^c_{i}(t)\}& =& \mu^c_{X}(t) + \sum_{k=1}^{K_{X,c}} \ \psi^c_{X,k} (t)  \xi^c_{X,ik},
\label{eqn:fittedbernoulli_model}
\eeqn 
where the functions $\mu^c_{X}(t)$ and $\psi^c_{X,k} (t) $ are replaced by their estimates and  $\xi^c_{X,ik}$ are uncorrelated and assumed to be normally distributed with zero mean and variance equal to $\widehat\lambda^c_{X,k}$. To eliminate possible bias \citep{gertheiss2017note}, we perform a two-step estimation where the class-specific mean function is re-estimated together with the prediction of the scores $\xi^c_{X,ik}$'s using model (\ref{eqn:fittedbernoulli_model}). Denote by $\widehat \xi^c_{X,ik}$ the predicted scores, for $k=1, \ldots, K_{X,c}$. The user-specific posting probability is estimated as $\widehat p^c_{i}(t) = g^{-1} \left \{ \widehat{X}_i^{c,K_{X}}(t)\right \} $, where $\widehat{X}_i^{c,K_{X}}(t)=\widehat  \mu^c_{X}(t) + \sum_{k=1}^{K_{X,c}} \ \widehat \psi^c_{X,k} (t)  \widehat \xi^c_{X,ik}, \label{eqn:X_rec}
$ for $t\in \mathcal{T}$, is the truncated KL reconstruction.

\subsubsection{Latent censored-data process under informative sampling}\label{ssecE2}
Recovering the latent process $Z_i(\cdot)$ underlying the censored observations is complicated by the informative sampling design. Building on FPCA methods for censored functional data under noninformative sampling \citep{dey2024functional,murphy2026}, we estimate the class-specific mean and covariance functions using inverse-probability weighting to account for informative observation times. We use a two-step procedure in which raw mean and covariance estimates are obtained via local weighted likelihood estimation and then smoothed. Unlike \citet{murphy2026}, estimation is performed separately within each class and incorporates inverse-probability weighting to account for informative sampling.

We denote by $p^c_{ij}=p^c_{i}(t_{ij})$ the probability that account $i$ in class $c$ posted at time $t_{ij}$; Section \ref{ssecE1} discussed estimation of the probability trajectories $p^c_{i}(t), t\in\mathcal{T}$,  separately for each group (class $c$). However, to reduce notational complexity, we omit the group-specific notation for the remainder of this subsection. Let $t_1, \ldots, t_m$ be an equally-spaced grid of points spanning the time domain $\myT$ and denote by $\mu_{Z,l}$ and $\sigma_{Z,ll'}$ the group-specific mean and the covariance of the underlying latent process $Z(\cdot)$ at times $t_l$ and $(t_{l}, t_{l'})$, respectively, where $l\neq l'$. When $t_l=t_{l'}$, the uncensored variance $\tau^2_{Z,l}$ involves both the variance of the latent process at $t_l$, $\sigma^2_{Z_{ll}}$, and the measurement error variance $\sigma^2_{Z,\epsilon}$, as $\tau^2_{Z,l} = \sigma^2_{Z_{ll}} + \sigma^2_{Z,\epsilon}$.

Step 1: Raw estimation of group-specific mean and variance. Fix class $c$ and suppress the dependence on $c$ in $p_{ij}$, $\mu_{Z,l}$, and $\tau_{Z,l}$. We estimate the class-specific mean $\mu_{Z,l}$ and standard deviation $\tau_{Z.l}$ using locally weighted maximum likelihood under a working independence Gaussian model. Let $\Phi(\cdot)$ and $\phi(\cdot)$ be the cdf and pdf, respectively, of $N(0,1)$. Define $\ell_{marg}(\mu_{Z,1}, \tau_{Z,1}^2\ldots, \mu_{Z,m}, \tau_{Z,m}^2) = \sum_{l=1}^m \ell(\mu_{Z,l}, \tau_{Z,l}^2) $, where $\ell(\mu_{Z,l}, \tau_{Z,l}^2)$ is equal to 
\beqn
\sum_{i: Y_i=c} \sum_{j=1}^{m_i} &
\widehat p_{ij}^{-1}  \left[   1\left (W_{ij} =a \right ) \log\left \{ \Phi\left( \frac{a-\mu_{Z,l}}{\tau_{Z,l}}\right) \right\}  +
1\left (W_{ij} =b \right) \log \left \{ 1- \Phi\left( \frac{b-\mu_{Z,l}}{\tau_{Z,l}} \right)\right\} \right] K_h(t_{ij}-t_l) \nonumber \\
 &+{\widehat p_{ij}}^{-1} 1\left (W_{ij} \in (a,b) \right) \log\left \{ \phi \left( \frac{W_{ij}-\mu_{Z,l}}{\tau_{Z,l}} \right)\right\}   K_h(t_{ij}-t_l);
\label{eqn:marginal_independence_likel_fn}
\eeqn 
where $K_h(t)$ is a kernel function with bandwidth kernel equal to $h$.
We use a uniform kernel, $K_h(t)=1(|t|\le h)$, and undersmooth the estimates to accommodate the subsequent smoothing step.

To estimate the covariance function, we use a weighted pairwise likelihood based on the previously estimated mean and variance functions. Let $\ell_{12}(\mu_1,\mu_2,\tau_1^2,\tau_2^2,\rho;W_1,W_2)$ denote the likelihood contribution of a generic bivariate censored observation $(W_1,W_2)$ under a Gaussian model with means $\mu_1,\mu_2$, variances $\tau_1^2,\tau_2^2$, and correlation $\rho$. Then $\ell_{12}$ equals
\beqn
    \begin{cases}
        \log\Phi_2 \left(\begin{pmatrix} a\\ a\end{pmatrix}; \begin{pmatrix} {\mu}_1\\ {\mu}_2\end{pmatrix}, \begin{pmatrix}
            \tau_1^2 & \rho \tau_1\tau_2\\
            \rho \tau_1\tau_2 &  \tau_2^2 
            \end{pmatrix}\right) & \text{ if } W_1=a, W_2=a\\

\log\Phi_C \left (a|W_1; \mu_1, \mu_2,\tau_1^2 ,\tau_2^2 ,\rho\right)+\log\phi(W_1; \mu_1, \tau_1^2 ) & \text{ if } W_1\in(a,b), W_2=a\\

\log\Phi_C\bigg(a|W_2=b; \mu_1, \mu_2,\tau_1^2, \tau_2^2, \rho \bigg)+\log(1-\Phi(b; \mu_2, \tau_2^2)) & \text{ if } W_1=a, W_2=b\\

\log\Phi_C\bigg(a|W_2; \mu_1, \mu_2,\tau_1^2,\tau_2^2,\rho\bigg)+\log\phi(W_2; \mu_2, \tau_2^2) & \text{ if } W_1=a, W_2\in(a,b)\\

 \log\phi_2 \left(\begin{pmatrix} W_1\\ W_2\end{pmatrix}; \begin{pmatrix} {\mu}_1\\ {\mu}_2\end{pmatrix}, \begin{pmatrix}
            \tau_1^2 & \rho \tau_1\tau_2\\
            \rho \tau_1\tau_2 &  \tau_2^2 
            \end{pmatrix}\right) & \text{ if } W_1\in(a,b), W_2\in(a,b)\\

\log(1-\Phi_C(b|W_2; \mu_1, \mu_2, \tau_1^2, \tau_2^2, \rho))+\log\phi(W_2; \mu_2, \tau_2^2) & \text{ if } W_1=b, W_2\in(a,b)\\

\log\Phi_C\bigg(a|W_1=b; \mu_1, \mu_2,\tau_1^2, \tau_2^2 \rho\bigg)+\log(1-\Phi(b; \mu_1, \tau_1^2)) & \text{ if } W_1=b, W_2=a\\

\log(1-\Phi_C(b|W_1; \mu_1, \mu_2,\tau_1^2, \tau_2^2, \rho))+\log\phi(W_1; \mu_1, \tau_1^2) & \text{ if } W_1\in(a,b), W_2=b\\

\log(1-\Phi_C(b|W_1=b; \mu_1. \mu_2, \tau_1^2, \tau_2^2, \rho))+\log(1-\Phi(b; \mu_1, \tau_1^2)) & \text{ if } W_1=b \ W_2=b \\
\end{cases}
\eeqn
where $\Phi_2(\cdot; \boldsymbol{\mu}, \boldsymbol{\Sigma})$ is bivariate Guassian cumulative distribution function with bivariate mean vector $\boldsymbol{\mu}$ and $2\times 2$ covariance matrix $\Sigma$, $\Phi_C(\cdot|W; \mu_1,\mu_2,\tau_1^2,\tau_2^2,\rho)$ is the conditional Gaussian cumulative distribution function, conditional on $W$, with parameters $\mu_1,\mu_2,\tau_1^2,\tau_2^2,\rho$, and $\phi(\cdot,\mu,\tau^2)$ is the Gaussian density function with mean $\mu$ and variance $\tau^2$.

Let the pairwise local likelihood function $\ell_{pair}(\rho_{12}, \rho_{13}, \ldots, \rho_{(m-1)m}) = \sum_{l<l'}^m \ell_{pair,ll'} (\rho_{ll'}) $, and   
\beqn
\ell_{pair,ll'} (\rho_{ll'})=\sum_{i:Y_i=c} \sum_{j<j'}^{m_i} 
\widehat p_{ij}^{-1}  \widehat p_{ij'}^{-1} \ell_{jj'} \left  (\widetilde \mu_{Z,l}, \widetilde\mu_{Z,l'}, \widetilde {\tau}_{Z,l}^2,\widetilde{\tau}_{Z,l'}^2, \rho_{ll'}; W_{ij}, W_{ij'} \right) K_h(t_{ij}-t_l)K_h(t_{ij'}-t_{l'}),
\eeqn 
where $\widetilde{\mu}_{Z,l}$ and $\widetilde {\tau}^2_{Z,l}$ are the raw estimates obtained in the initial stage. 

Step 2: Final estimation of the class-specific mean and covariance functions. Fix class $c$. The raw estimates from Step 1 are smoothed using penalized spline methods \citep{ruppert2003semiparametric,wood2000modelling,wood2016smoothing}. Representing $
\mu_Z(t)=\sum_{r=1}^R B_r(t)\beta_r,$
the coefficients are estimated by penalized least squares with smoothing parameter selected by restricted maximum likelihood (REML), yielding
$\widehat{\mu}_Z^c(t)=\sum_{r=1}^R B_r(t)\widehat{\beta}^c_r$.
%
%
Estimation of $\Sigma_Z(\cdot,\cdot)$ requires accounting for measurement error. Following standard functional data analysis methodology, a bivariate smoother is applied to the off-diagonal covariance estimates, and the diagonal is subsequently recovered by prediction. Full details are provided in the Supplementary Material. Let $\widehat{\Sigma}_Z^c(t,t')$ denote the resulting smooth covariance estimate for class $c$.


Let$\{\widehat{\lambda}^c_{Z,k}, \widehat{\psi}^c_{Z,k}(\cdot)\}_k$ be the eigencomponents of $\widehat{\Sigma}^c_{Z}(t,t')$, and let $K_{Z,c}$ denote a truncation level. The latent trajectories $Z^c_i(\cdot)$ are recovered using the truncated KL expansion
\beqn
{Z}^{c,K_{Z}}_i(t) = {\mu}^c_{Z}(t) + \sum_{k=1}^{K_{Z,c}} {\psi}^c_{Z,k}(t){\xi}^c_{Z,ik}, \label{eqn:Z_rec}
\eeqn
where ${\xi}^c_{Z,ik}$ are uncorrelated across indices $k$, have mean zero and variance ${\lambda}_{Z,c,k}$. The scores ${\xi}^c_{Z,ik}$ are predicted using conditional expectation, $\widetilde{\xi}^c_{Z,ik} = E[\xi^c_{Z,ik} \mid \mathbf{W}_i, Y_i=c]$, where $\mathbf{W}_i = (W_{i1}, \ldots, W_{im_i})^T$, and replacing ${\mu}^c_{Z}(t)$, ${\psi}^c_{Z,k}(t)$, and ${\lambda}^c_{Z,k}$ by their estimates. Due to the censored data $\mathbf{W}_i$, the conditional expectation does not have an analytical expression \citep{yao2005}; 
however we can use Monte Carlo simulations and a working normal distribution for the scores to estimate this conditional distribution. Specifically, let $\bxi^c_{Z,i} = (\xi^c_{Z,i1}, \ldots, \xi^c_{Z,iK_{Z,c}})^T$ , and assume that $\xi^c_{Z,ik} $ are normally distributed with mean zero and variance equal to ${\lambda}^c_{Z,k}$ and are mutually uncorrelated. Using Bayes rule we have
\beqn 
    \mathbb{E} \left [\bxi^c_{Z,i}|\bW_{i}, Y_i=c\right ] &=&  \frac{  \int   \bxi^c_{Z,i}f_{\bW}(\bW_{i}|\bxi^c_{Z,i}) \times \pi_{Z,c}(\bxi^c_{Z,i}) \  d\bxi^c_{Z,i} } {\int f_{\bW}(\bW_{i}|\bxi^c_{Z,i}) \times \pi_{Z,c}(\bxi^c_{Z,i}) \
    d\bxi^c_{Z,i} }  \nonumber \\
    &=&  \frac{  \int   \bxi^c_{Z,i}  \prod_{j=1}^{m_i} f_{W_{ij} }(W_{ij}|\bxi^c_{Z,i}) \times \pi_{Z,c}(\bxi^c_{Z,i}) \  d\bxi^c_{Z,i} } {\int \prod_{j=1}^{m_i} f_{W_{ij} }(W_{ij}|\bxi^c_{Z,i}) \times \pi_{Z,c}(\bxi^c_{Z,i}) \
    d\bxi^c_{Z,i} },
    \label{eqn:predicting_scores_censored_FD}
\eeqn 
where $\pi_{Z,c} (\boldsymbol{\xi})$ is the density of a $K_{Z,c}$- dimensional multivariate normal with zero mean and covariance equal to the diagonal matrix $diag\{\lambda^c_{Z,1}, \ldots, \lambda^c_{Z,K_{Z,c}} \}$.
We have $f_{W_{ij}}(W_{ij}|\bxi^c_{Z,i})=f_{ij}\left(W_{ij}; \mu^c_{Z}(t_{ij})+\sum_{k=1}^{K_{Z,c}} \xi^c_{Z,ik}\psi^c_{Z,k}(t_{ij}),\sigma^2_{Z,\epsilon}\right)$, where $f_{ij}(\cdot, \mu_{ij}, \sigma^2)$ denotes the density of a normal distribution censored to the interval
$[a,b]$, with mean $\mu_{ij}$ and variance $\sigma^2$. In particular, denote by $\mu^{c*}_{Z,ij}=\mu^c_{Z}(t_{ij})+\sum_{k=1}^{K_{Z,c}} \xi^c_{Z,ik}\psi^c_{Z,k}(t_{ij})$, then $f_{W_{ij}}(W_{ij}; \mu^{c*}_{Z,ij},\sigma^2_{Z,\epsilon})$ equals
\beqn
& \exp\bigg\{\left[   1\left (W_{ij} =a \right ) \log\left \{ \Phi\left( \frac{a-\mu^{c*}_{Z,ij}}{\sigma_{Z,\epsilon}}\right) \right\}  +
1\left (W_{ij} =b \right) \log \left \{ 1- \Phi\left( \frac{b-\mu^{c*}_{Z,ij}}{\sigma_{Z,\epsilon}} \right)\right\} \right]\nonumber \\
 &+1\left (W_{ij} \in (a,b) \right) \log\left \{ \phi \left( \frac{W_{ij}-\mu^{c*}_{Z,ij}}{\sigma_{Z,\epsilon}} \right)\right\} \bigg\}.
 \label{eqn:censored_density}
\eeqn 
Let $\widehat{f}_{W_{ij}}(W_{ij}|\bxi^c_{Z,i})$  be $f_{W_{ij}}(\cdot| \bxi^c_{Z,i})$ with $\mu^c_{Z}$, $\psi^c_{Z,k}$, $\lambda^c_{Z,k}$, and $\sigma^2_{Z,\epsilon}$ substituted by their estimated versions,
$\widehat \mu^c_{Z}$, $\widehat \psi^c_{Z,k}$, $\widehat \lambda_{z,c,k}$, and $\widehat \sigma^2_{Z,\epsilon}$, respectively. The scores ${\bxi}_{Z,c,i}$ can thus be predicted by 
\beqn \widehat{\bxi}^c_{Z,i}=\frac{\sum_{s=1}^{S}\bxi^c_{s}\prod_{j=1}^{m_i}\widehat{f}_{W_{ij}}(W_{ij}|\bxi^c_{s})}{\sum_{s=1}^{S}\prod_{j=1}^{m_i}\widehat{f}_{W_{ij}}(W_{ij}|\bxi^c_{s})} \label{eqn:xiZl}
\eeqn 
for large $S$, where $\bxi^c_{s}=(\xi^c_{s,1},\ldots,\xi^c_{s,K_{Z,c}})^T$ are independent Monte Carlo draws from $\widehat \pi_{Z,c} $ - like $\pi_{Z,c} $ described above, except $ \lambda^c_{Z,k}$'s are replaced by $ \widehat \lambda^c_{Z,k}$, $k=1, \ldots, K_{Z,c}$.
While computationally more efficient than the approach of \cite{murphy2026}, this method accounts only for censoring and not the informative sampling design. We therefore introduce a feature extraction procedure that accommodates both sources of complexity.

\subsubsection{Low-dimensional feature extraction}\label{ssecE3}

After estimating the latent processes for the censored observations and sampling design, we extract low-dimensional features via multivariate functional principal component analysis. Let $\boldsymbol{\mu}(t)$ and $\boldsymbol{\Sigma}(t,t')$ denote the marginal mean and covariance functions of $(Z_i(\cdot),X_i(\cdot))^T$, with eigenpairs $\{(\lambda_k,\boldsymbol{\phi}_k)\}_{k\ge1}$. The multivariate KL expansion is
%
%
%
\begin{eqnarray}
   \begin{pmatrix}
        Z_i(t)\\ X_i(t)
   \end{pmatrix}=\boldsymbol{\mu}(t)+\sum_{k\geq{1}}\zeta_{ik}\boldsymbol{\phi}_k(t) \text{ for } t\in\mathcal{T}, i=1,\ldots,n \label{eqn:MFPCA_form}
\end{eqnarray}
where $\zeta_{ik}=\langle \langle (Z_i,X_i)^T-\boldsymbol{\mu}, \  \boldsymbol{\phi}_k\rangle\rangle_{\mathcal{L}^2(\mathcal{T})\times\mathcal{L}^2(\mathcal{T})}$ are functional principal component (FPC) scores with zero mean and variance equal to $\lambda_k$. Here $\langle \langle \cdot, \cdot \rangle \rangle$ is the usual inner product in $\mathcal{L}^2(\mathcal{T})\times\mathcal{L}^2(\mathcal{T})$. We use the resulting score vectors as a low-dimensional feature representation.

We obtain marginal FPC scores by applying \citet{pomann2016two} separately to $Z_i(\cdot)$ and $X_i(\cdot)$, and then perform the multivariate KL decomposition (\ref{eqn:MFPCA_form}) using the approach of \citet{HappGreven2018} with a novel scaling step. The scaling prevents components with disproportionately large variability from dominating the decomposition.


The marginal mean and covariance functions are obtained directly from their class-specific counterparts. For example,
$\mu_X(t) =p \mu^1_{X}(t)+ (1-p)\mu^0_{X}(t)$ and the covariance is 
$\Sigma_X(t,t')= p \Sigma^1_{X}(t,t')+ (1-p)\Sigma^0_{X}(t,t') +  p(1-p)\left \{ \mu^1_{X}(t)-\mu^0_{X}(t) \right \} \left \{\mu^1_{X}(t')-\mu^0_{X}(t')\right \}$, where $p$ represents the probability of class $1$ and $(1-p)$ the probability of class $0$. Let $\{\lambda_{X,k},\psi_{X,k}(\cdot)\}_{k\geq{1}}$ denote the eigenpairs of $\Sigma_X$. A truncated KL expansion then yields a low-dimensional representation of $X_i(\cdot)$, say $\widehat X_i(\cdot) = \widehat \mu_X(t)  + \sum_{k=1}^{K_X} \widehat \psi_{X,k}(t) \xi^X_{ik}$, with scores estimated by projection of $\widehat X_i^{K_{X,c}}(\cdot)$ onto the marginal eigenfunctions, $\widehat{\xi}^X_{ik} = \int \{ \widehat X^{K_{X,c}}_i(t) -\widehat \mu_X(t)\} \widehat \psi_{X, k}(t) \, dt$. An analogous construction is used for $Z_i(\cdot)$, say $ \widehat Z_i(t) =\widehat \mu_Z(t)  + \sum_{l=1}^{K_Z} \widehat \psi_{Z,l}(t) \widehat  \xi^Z_{il}$.

Let $\boldsymbol{\Psi}(t)$ denote the block matrix of marginal eigenfunctions for $Z_i(\cdot)$ and $X_i(\cdot)$, with truncation levels chosen to satisfy a prespecified proportion of variance explained (PVE). Common features can be extracted using the multivariate KL decomposition of \citet{HappGreven2018}, which rotates $\boldsymbol{\Psi}(t)$ according to the covariance structure of the marginal FPC scores. To prevent high-variance components from dominating the decomposition, we first standardize each score to have unit variance and then perform eigenanalysis of the correlation matrix of the pooled standardized scores. Specifically if $s\boldsymbol{\xi}_i$ denotes the vector of the standardized scores from $Z_i(\cdot)$ and $X_i(\cdot)$, with entries $\xi^Z_{ik}/\sqrt{\lambda_{Z,k}}$ 
and $\xi^X_{il}/\sqrt{\lambda_{X,l}}$, 
then
$
\Gamma=\mathrm{Cor}(s\boldsymbol{\xi}_i)=UDU^T$.
The leading eigenvectors in $U_K$ define the shared features across the two processes, with $K$ chosen to explain a prespecified proportion of variation.

Using the selected truncation, we approximate the covariance operator by
$
\widetilde{\Sigma}(t,t')
=
\boldsymbol{\Psi}(t)\Lambda^{1/2}U_KU_K^T\Lambda^{1/2}\boldsymbol{\Psi}(t')^T,
$
where $\Lambda$ is the $(K_Z+K_X)$ diagonal matrix with elements $\lambda_{Z,k}$ and $\lambda_{X,l}$.  Let $\{\boldsymbol{\phi}_k(\cdot)\}_{k=1}^K$ denote the resulting eigenfunctions. In practice, $\Gamma$ is estimated from the standardized score vectors, and the multivariate features $\widehat{\zeta}_{ik}$ (\ref{eqn:MFPCA_form}) are obtained by projecting the centered bivariate curves onto the estimated directions, $\widehat {\boldsymbol{\phi}}_k(\cdot)$.

For a new account with posting data $\boldsymbol{W}_{new}$ and observation indicators $\boldsymbol{R}_{new}$, the features are obtained by projecting the predicted latent processes $\widehat Z_{new}(\cdot)$ and $\widehat X_{new}(\cdot)$ onto the estimated multivariate eigenfunctions,
$
\widehat{\zeta}_{new,k}
=
\left\langle\!\left\langle
(\widehat Z_{new},\widehat X_{new})^T-\widehat{\boldsymbol{\mu}},
\widehat{\boldsymbol{\phi}}_k
\right\rangle\!\right\rangle,
$
where $\widehat Z_{new}(t)=\widehat E\{Z_{new}(t)\mid \boldsymbol{W}_{new}\}$ and $\widehat X_{new}(t)=\widehat E\{X_{new}(t)\mid \boldsymbol{R}_{new}\}$. Furthermore, we calculate
\beqn
\widehat{E}[Z_{new}(t)\, \big  |\boldsymbol{W}_{new}] = 
\pi \widehat{E}[Z_{new}(t)\, \big  |\boldsymbol{W}_{new}, Y_{new}=1] + (1-\pi) \widehat{E}[Z_{new}(t)\, \big  |\boldsymbol{W}_{new}, Y_{new}=0];
\nonumber
\eeqn 
where $\pi=P(Y_i=1)$ is the prevalence of class $1$. Here $\widehat{E}[Z_{new}(t)\, \big  |\boldsymbol{W}_{new}, Y_{new}=c] = \widehat \mu^c_{Z}(t) + \sum_{k=1}^{K_{Z,c}} \widehat \psi^c_{Z,k}(t) \ \widehat{E} [\xi^c_{Z, new \ k}\, | \boldsymbol{W}_{new}, Y_{new}=c]$ using the KL representation (\ref{eqn:Z_rec}) described for class $c$. Estimation of $E[X_{new}(t)|\boldsymbol{R}_{new}]$ is done similarly  (see Section \ref{eqn:X_rec}).

\subsubsection{Estimation of the predictive model}\label{ssecE4}

Once low-dimensional user-specific features $\widehat{\boldsymbol{\zeta}}_i$ are extracted, prediction of user status $Y_i$,
follows directly from literature of parametric and non-parametric regression. In particular, for a linear predictor
$
h\{P(Y_i = 1|\boldsymbol{\zeta}_i)\} = \beta_0 + \boldsymbol{\beta}^T \boldsymbol{\zeta}_i,
$
the parameters $\beta_0$ and $\boldsymbol{\beta}$ can be estimated by maximum likelihood under a Bernoulli model (see \cite{mccullagh1989generalized}). This framework allows for more flexible predictors, such as additive, $h\{P(Y_i = 1|\boldsymbol{\zeta}_i)\}  = f^*_{1}(\zeta_1) + \ldots + f^*_{K}(\zeta_{K})$, for unknown univariate smooths $f^*_k(\cdot)$'s, or non-additive $h\{P(Y_i = 1|\bzeta_i)\} = f^*(\bzeta_i)$ for unknown multivariate smooth $f^*(\cdot)$ -  the unknown functions are estimated using the penalized negative log-likelihood; see \cite{gam1986, eilersMarx1996, wood2006} etc.

These models are well established, so we provide only a brief overview. Unknown functions are represented using basis expansions and estimated via penalized smoothing. We consider both isotropic thin-plate spline smooths and non-isotropic tensor-product smooths, with smoothing parameters selected by generalized cross validation (GCV) or REML \citep{wood2011}. All models are fitted using the \texttt{R} package \texttt{mgcv} \citep{wood2006generalized}. For both the simulation study and data analysis, we use a logistic link, $h(\cdot)=logit(\cdot)$. The fitted model is used for out-of-sample prediction. We refer to the proposed methodology as \emph{Scalar-on Censored Informative-design Functional Regression} (SoCIFR).

\section{Extensions}\label{sec:extensions}

We consider extensions to matched case-control designs and to multiple functional predictors subject to censoring and/or informative missingness.


\subsection{
Matched scalar-on censored informative-design functional regression} \label{ssec_Meth2}

In matched case-control studies, such as the motivating application, ignoring the matching design and applying the logistic regression model in Section~\ref{ssec_Meth1} may lead to biased estimation \citep{breslow1980statistical}, particularly when the relationship between the matching factors and the outcome, after adjustment for the main covariates, is unknown \citep{Wan2021}.
To clarify, consider the \textcolor{black}{1:1 match-pair case-control} data, where each pair $i$ consists of $[\{({W}_{1ij},t_{1ij})_{j=1}^{m_{i1}}, Y_{i1}\}, 
\{({W}_{2ij},t_{2ij})_{j=1}^{m_{i2}}, Y_{i2}\}, \boldsymbol{M}_i]$ 
\textcolor{black}{where $Y_{i1},Y_{i2}\in\{0,1\}$, such that $Y_{i1}+Y_{i2}=1$},
for all $i=1, \ldots, n$, $(W_{1ij}, t_{1ij})$'s are censored FD observed on informative design, and $\boldsymbol{M}_i$ are matching factors. Under a logit link and a linear predictor within each pair, we have
\beqn
    logit \{ P (Y_{ip}=1| Z_{ip}, X_{ip})\} = \beta_{i,0} + \boldsymbol{\gamma}^T\boldsymbol{M}_i+
    \int Z_{ip}(t)\beta_{Z}(t) dt +\int X_{ip}(t)\beta_{X}(t) dt \label{eqn:paired_logit}
\eeqn
where \(p \in \{1,2\}\) indexes the members of pair \(i\), \(\beta_{i,0}\) is a pair-specific intercept treated as a nuisance parameter, $\boldsymbol{\gamma}$ is the effect of the matching factors,
and the parameter functions $\beta_Z(\cdot)$ and $\beta_X(\cdot)$ are smooth, unknown functions of primary interest. Because the number of pair-specific nuisance parameters increases with the sample size, direct likelihood-based estimation suffers from the incidental parameters problem \citep{neyman1948consistent}. We base inference on a conditional likelihood that eliminates these nuisance parameters, following the classical framework of conditional logistic regression for matched case–control studies (\cite{breslow1980statistical, agresti2013, Keogh2014}).

As described in Section~\ref{sec:framework}, we first extract the latent processes \(Z_{ip}(\cdot)\) and \(X_{ip}(\cdot)\), \(p=1,2\), under a working independence assumption within pairs, and then combine them to obtain the common features \(\bzeta_{ip}\). In our data application, the correlation between the processes within matched pairs is negligible, so an analysis based on a working independence assumption is reasonable. When there is evidence of within-pair correlation, however, a different approach would be needed; we leave this extension for future work. 
Let $\boldsymbol{\zeta}_{i1},\boldsymbol{\zeta}_{i2}$ be the low-dimensional features from the censored functional covariates from the $i$th matched pair 
 $(Z_{ip}(\cdot), X_{ip}(\cdot))^T$ for $p=1,2$.  The conditional probability can be derived from model (\ref{eqn:paired_logit}) assuming conditional independence of the responses within the pair given the low-dimensional features;
\beqn
P(Y_{i1}=1, Y_{i2}=0|\boldsymbol{\zeta}_{i1}, \boldsymbol{\zeta}_{i2}, Y_{i1}+Y_{i2}=1) = \exp(\boldsymbol{\beta}^T \bzeta_{i1})/\{\exp(\boldsymbol{\beta}^T \bzeta_{i1})+\exp(\boldsymbol{\beta}^T \bzeta_{i2})\}, \label{eqn:condLog}
\eeqn
this is free of the nuisance parameter $\beta_{i,0}$. 

While model (\ref{eqn:condLog}) can be used to predict/classify discordant pairs, we can also use it to predict user status $Y_{ip}$ via either unadjusted or adjusted conditional logistic regression. For this, we adopt a two-stage parameter estimation methodology \citep{Qian2014, Xu2019}. In the first stage, we use conditional logistic regression (\ref{eqn:condLog}) to estimate the regression parameter $\boldsymbol{\beta}$; denote this estimate by $\widehat{\boldsymbol{\beta}}^{CLR}$. In the second stage, we fit an unconditional model to estimate the population-level intercept and the effects of the matching factors, with \((\widehat{\boldsymbol{\beta}}^{\mathrm{CLR}})^T \boldsymbol{\zeta}_{ip}\) included as a fixed linear predictor. Specifically, we use the following approximate model to predict the outcome for an individual subject, rather than the outcome assignment within a matched pair:
\beqn
logit \{ P(Y_{ip}=1 \mid \boldsymbol{\zeta}_{ip}) \}
=
\beta_0 + \boldsymbol{\gamma}^T \boldsymbol{M}_i
+ (\widehat{\boldsymbol{\beta}}^{\mathrm{CLR}})^T \boldsymbol{\zeta}_{ip}.
\label{eqn:adj_clr}
\eeqn
Prediction for a new account with matching covariates \(\boldsymbol{M}_{{new}}\) can be obtained from model~(\ref{eqn:adj_clr}) by plugging in the corresponding parameter estimates and computing $\bzeta_{new}$ as described in Section \ref{ssecE4}. Moreover, prediction for a new account can be adjusted to reflect the true population prevalence of cases; under the \(1{:}1\) matched design used for estimation, the implied case prevalence is \(50\%\); see \cite{King_Zeng_2001}.

\subsection{ Extension to Multiple Functional Covariates with Censoring and/or Missingness} \label{ssec_Meth3}

The proposed methodology extends naturally to settings with multiple functional covariates, some subject to censoring and/or informative sampling and others observed under noninformative designs. For feature extraction, covariates are grouped according to their sampling mechanisms. 

Suppose subject \(i\) has \(d_1\) functional covariates observed under a common informative sampling design, \(\boldsymbol{W}_i^{1,d}\), \(d=1,\ldots,d_1\), and \(d_2\) covariates observed under a noninformative design, \(\boldsymbol{W}_i^{2,d}\), \(d=1,\ldots,d_2\). Feature extraction is performed separately within each group. The noninformative covariates are analyzed jointly using MFPCA with the proposed scaling. For the informative group, we jointly recover the latent covariate processes and the latent observation process, and then apply MFPCA with the proposed scaling to obtain shared low-dimensional features.For the first group, we jointly recover the latent processes underlying the observed covariates, say $(\widehat{Z}_i^{1,1}(\cdot),\ldots,\widehat{Z}_i^{1,d_1}(\cdot))$, together with the latent process associated with the shared informative sampling design; say $\widehat X_i(\cdot)$. These latent processes are modeled jointly using MFPCA to extract low-dimensional features, as in Section~\ref{ssecE2}.

This setting arises in our data application, where the functional covariates are of mixed type: weekly average post volume is uncensored and observed on a regular noninformative sampling design, whereas average capped \( @ \)-tweet rate, taking values in \([0,1]\), and average post length, which is strictly positive, are both observed under an informative time design.

\section{Numerical Simulations}\label{sec:simulations}

\subsection{Regression with Censored Functional Covariates under Informative Sampling}\label{sssec:simA}

We consider two simulation designs for the covariates. Scenario A involves one censored functional covariate observed under an informative observation design, \(\{(W_{ij}, t_{ij}): j=1,\ldots,m_i\}_i\). Scenario B extends this setting to two censored functional covariates observed under a common informative design, \(\{(W_{1,ij}, W_{2,ij}, t_{ij}): j=1,\ldots,m_i\}_i\). In both scenarios, the latent processes are generated using finite-dimensional multivariate functional principal component models. Observation times are then sampled through a Bernoulli mechanism that depends on the latent design process, inducing informative observation patterns. Finally, the binary outcome is generated from a logistic regression model based on the latent subject-specific scores, with different signal levels used to control the prevalence of positive outcomes. Full details of the data-generating mechanisms, including the mean functions, eigenfunctions, eigenvalues, measurement error variances, and the values of \(\beta_0\) and \(\boldsymbol{\beta}\), are provided in Section S.2 of the Supplementary Material. We evaluate recovery of the mean and eigenfunctions using integrated squared error (ISE), subject-specific scores using MSE, and functional covariates using mean integrated squared error (MISE). Overall, the functional covariates are recovered with increasing accuracy as the sample size grows. For brevity, the results are reported in Tables S1 and S2 of the Supplementary Material

Prediction performance is evaluated on an independent test set of size \(n_{\mathrm{test}}=500\). We fit the proposed SoCIFR model using a logistic link with linear (M1), additive (M2), and nonlinear (M3) effects of the MFPCA scores. Since the data-generating mechanism is linear, M2 and M3 perform similarly to M1; these results are reported in the Supplementary Material. In the data application, M2 and M3 are included to allow for more flexible predictor effects.

We compare the predictive performance of \emph{SoCIFR} with two random forest alternatives designed to account for the informative sampling and longitudinal structure of the data. The first approach (A1) is a \textit{naive} random forest implementation that preserves the temporal ordering of the functional covariates by using the observed values in their measurement order and treating unobserved time points as missing. Missing values are subsequently imputed using the median value of the corresponding covariate at that time point across the remaining users. This approach exploits the regular underlying grid and is therefore not directly generalizable to irregular designs. The second approach (A2) uses as inputs the low-dimensional subject-specific features extracted by the proposed method described in SoCIFR; we refer to it as the \textit{feature-based} random forest. Both random forest models are fitted using the \texttt{randomForest} package in \textsf{R} \citep{randomForest}. We use the default tuning parameters, with 500 trees and \(\sqrt{p}\) candidate variables sampled at each split, where \(p\) denotes the number of input predictors.

Table~\ref{tab:tabMain1} compares the mean PR AUC for SoCIFR (M1) with the two random forest variants (A1 and A2), across 100 Monte Carlo simulations for varying training sample sizes (\(n\)) and signal levels (\(\delta\)) that control the prevalence of the cases in the training set ($\pi$). A larger $\delta$ value is associated with a larger prevalence $\pi$ and with sharpness in the boundary decision rule. This, in turn, leads to increased predictive performance overall. The proposed SoCIFR has a competitive predictive performance to the random forests based variants. Similar results are observed for Scenario B; to avoid repetition, we include the results in Table S5 of the Supplementary Material.

\begin{table}[H]
\centering
    \begin{tabular}{|c|c|c|c|c|c|}
\hline
$\delta$ & n & $\pi$ & Naive RF (A1) & Feature-based RF (A2) & SoCIFR (M1)\\
\hline
0 & 200 & 0.50 (0.04) & 0.50 (0.03) & 0.50 (0.03) & 0.51 (0.03)\\

 & 500 & 0.50 (0.02) & 0.50 (0.03) & 0.51 (0.03) & 0.50 (0.03)\\

 & 1000 & 0.50 (0.02) & 0.50 (0.03) & 0.50 (0.03) & 0.50 (0.03)\\
\hline
0.5 & 200 & 0.51 (0.03) & 0.55 (0.04) & 0.58 (0.03) & 0.64 (0.03)\\

 & 500 & 0.50 (0.02) & 0.56 (0.03) & 0.59 (0.03) & 0.65 (0.03)\\

 & 1000 & 0.50 (0.02) & 0.58 (0.03) & 0.60 (0.03) & 0.65 (0.03)\\
\hline
1 & 200 & 0.52 (0.03) & 0.62 (0.04) & 0.67 (0.04) & 0.73 (0.03)\\

 & 500 & 0.52 (0.02) & 0.65 (0.04) & 0.68 (0.03) & 0.74 (0.03)\\

 & 1000 & 0.52 (0.02) & 0.67 (0.03) & 0.69 (0.04) & 0.74 (0.03)\\
\hline
1.5 & 200 & 0.70 (0.03) & 0.79 (0.03) & 0.83 (0.03) & 0.87 (0.03)\\

 & 500 & 0.70 (0.02) & 0.81 (0.02) & 0.84 (0.02) & 0.88 (0.02)\\

 & 1000 & 0.70 (0.01) & 0.83 (0.02) & 0.86 (0.02) & 0.89 (0.02)\\
\hline
2 & 200 & 0.73 (0.03) & 0.81 (0.03) & 0.85 (0.03) & 0.89 (0.03)\\

 & 500 & 0.73 (0.02) & 0.84 (0.03) & 0.87 (0.02) & 0.90 (0.02)\\

 & 1000 & 0.73 (0.01) & 0.86 (0.02) & 0.88 (0.02) & 0.91 (0.01)\\
\hline
\end{tabular}
    \caption{Mean (SD) out-of-sample PR AUCs for Scenario A over 100 Monte Carlo simulations, by sample size \(n\) and signal level \(\delta\) (controlling outcome prevalence \(\pi\)), for SoCIFR (M1) and random forest approaches A1--A2.}
    \label{tab:tabMain1}
\end{table}

\subsection{Logistic Regression with Censored and Informative sampled Functional Covariates for Matched Case-Control Studies}\label{sssec:simB}

Simulation of matched case-control data requires a modified sampling design. In Scenario C, let \(i=1,\ldots,n/2\) index matched pairs and \(p=1,2\) the subjects within a pair. We first generate a large population from an unconditional logistic model in which the linear predictor depends on the latent functional data and a matching variable \(M_i\), with low outcome prevalence. Functional covariates are generated from model~(\ref{eqn_modelW}) with \(a=0\), \(b=1\), and measurement error variance \(1/16\). As in Scenarios A and B, we consider multiple signal levels indexed by \(\delta\); additional details are provided in Section~S.2 of the Supplementary Material. The matching variable \(M_i\) is sampled from a discrete uniform distribution with 300 levels to induce sparse stratification, and the latent processes use the same mean functions and eigenfunctions as in Scenario B.

Matched pairs \(\{(W_{p,ij}, t_{p,ij})_{j=1}^{m_{pi}}, M_i, Y_{ip}\}_{p=1,2}\), where \(Y_{ip}\in\{0,1\}\), are formed by selecting the first \(n/2\) cases and matching each to a control with the same value of \(M_i\). The corresponding functional and covariate information is then retained for both subjects in the pair.

We evaluate estimation accuracy for the parameter of interest \(\boldsymbol{\beta}\) using SoCIFR under conditional logistic regression, together with out-of-sample predictive performance measured by the area under the ROC curve (AUC). Results for recovery of \(\boldsymbol{\beta}\) are provided in the Supplementary Material. Here, we focus on predictive performance for both pair-level prediction using the conditional model~(\ref{eqn:condLog}) and subject-level prediction using model~(\ref{eqn:adj_clr}), across training sample sizes \(n\) ranging from 200 to 1000 and signal levels \(\delta\in\{0,0.5,1,1.5,2\}\). Table~\ref{tab:tabMain2} reports the mean out-of-sample AUC and corresponding standard deviation across 100 Monte Carlo simulations for pair-level prediction. The results show that predictive accuracy improves as the signal strength increases. Table~\ref{tab:tabMain3} presents the corresponding subject-level prediction results, which exhibit similar behavior; additionally, prediction accuracy improves modestly with increasing sample size.

\begin{table}[H]
\centering
\begin{tabular}{|c|c|c|c|c|c|}
  \hline
 & $\delta=0$ & $\delta=0.5$ & $\delta=1$ & $\delta=1.5$ & $\delta=2$ \\ 
  \hline
n=200  &  0.50 (0.03) & 0.67 (0.04) & 0.82 (0.03) & 0.88 (0.02) & 0.91 (0.02)  \\ \hline
n=500  &  0.50 (0.04) & 0.68 (0.03) & 0.82 (0.03) & 0.89 (0.02) & 0.92 (0.02)  \\ \hline
n=1000  &  0.50 (0.04) & 0.68 (0.03) & 0.82 (0.03) & 0.89 (0.02) & 0.92 (0.02)  \\ \hline
\end{tabular}
    \caption{Mean AUC (standard deviation) in 100 Monte Carlo simulations to predict pairs for various sample sizes $n$ and signal strength controlled by $\delta$.}
    \label{tab:tabMain2}
\end{table}

    \begin{table}[H]
    \centering
\begin{tabular}{|c|c|c|c|c|c|}
  \hline
 & $\delta=0$ & $\delta=0.5$ & $\delta=1$ & $\delta=1.5$ & $\delta=2$ \\ 
  \hline
n=200  &  0.50 (0.02) & 0.60 (0.02) & 0.70 (0.02) & 0.76 (0.02) & 0.78 (0.03)  \\ \hline
n=500  &  0.50 (0.02) & 0.61 (0.02) & 0.71 (0.02) & 0.77 (0.02) & 0.80 (0.02)  \\ \hline
n=1000  &  0.50 (0.02) & 0.62 (0.02) & 0.72 (0.02) & 0.78 (0.02) & 0.81 (0.02)  \\ \hline
\end{tabular}
    \caption{Mean AUC (standard deviation) in 100 Monte Carlo simulations to predict subjects for various sample sizes $n$ and signal strength controlled by $\delta$.}
    \label{tab:tabMain3}
    \end{table}

\section{Predicting Bot Vulnerability from Posting Trajectories}\label{sec:dataApp}

We now turn to our motivating application of predicting susceptibility to bot interaction and identifying the posting behaviors associated with increased susceptibility, using information observed during the preceding weeks. Recall that accounts are organized into matched pairs, consisting of one account that interacts with a bot and one account that does not. For each account, posting activity is observed over one year (52 weeks). For accounts that interact with bots, the first bot interaction occurs in the week immediately following this one-year observation window. In contrast, control accounts do not interact with bots for multiple years after the observation period. The data for account pair $i$ is $(D_{i1},D_{i2}; \textbf{M}_i)$, for $p=1,2$ with $D_{ip}=\Big\{ ( W_{p,ij}, V_{p,ij}, t_{p,ij})_{j=1}^{m_{ip}},\; (U_{p,il}, t_{l})_{l=1}^{52},\; Y_{ip} \Big\},$ where $(W_{p,ij}, V_{p,ij}, t_{p,ij})_{j=1}^{m_{ip}}$ records the average capped @-tweets rate ($W_{p,ij}$) and the average post length ($V_{p,ij}$) for each active week $t_{p,ij}$ during the one-year interval, and $m_{ip}$ denotes the total number of active weeks for user $p$ in pair $i$. Also, \((U_{p,il}, t_l)_{l=1}^{52}\) denotes the weekly average log-transformed post volume \(U_{p,il}\) recorded at week \(t_l \in \{1,2,\ldots,52\}\) over the 52-week study period. The outcome \(Y_{ip}\) indicates whether user \(p\) in pair \(i\) interacts with a bot (\(Y_{ip}=1\), ``susceptible'' user) or does not interact with a bot (\(Y_{ip}=0\), ``control'' user). The accounts are matched based on $\textbf{M}_i$, including the account reported creation year and US state. The goal is twofold. First, we aim to predict the probability that an account, whose posting behavior has been observed over one year, interacts with a bot in the subsequent week. Second, we aim to estimate the probability that an account interacts with a bot within a specified future time horizon, after observing its posting behavior over a given observation window.

Exploratory analysis indicates that Kendall's tau correlation between repeated observations from accounts within the same pair is negligible, based on the full dataset of 9,586 pairs; see Figure~S2 in the Supplementary Material. This likely reflects the relatively weak influence of the matching variables, account creation year and reported state, on account behavior and writing patterns. Thus, for both goals, the working independence assumption appears reasonable. Alternative approaches would be needed in applications where within-pair dependence is non-negligible.

The average capped @-tweet rate, $W_{p,ij}\in[0,1]$, is treated as an interval-censored functional covariate observed under an informative sampling design, while the average post length, $V_{p,ij}$, is treated as an uncensored, unbounded functional covariate observed under the same informative sampling design. In contrast, the average post volume, $U_{p,il}\in[0,\infty)$, is treated as regularly observed functional data, subject to left-censoring at the lower bound of $0$. To address the first goal, we use the SoCIFR methodology described in Section~\ref{sec:framework}, with the adjustments discussed in Section \ref{ssec_Meth2} and \ref{ssec_Meth3}, to accommodate multiple covariates of various types and the matched case control design.   

Feature extraction used 99\% and 95\% PVE thresholds in the univariate FPCA and correlation-matrix decomposition steps, respectively, yielding approximately 4--5 features per trajectory. These features were included in a conditional logistic regression model and evaluated using 10 random train--test splits (90\% training about 7,316 pairs, 10\% testing and 813 pairs). The estimated effects indicate that increased posting activity, longer tweets, and greater tweet volume become associated with elevated bot-interaction risk approximately 6--8 weeks before the reference week. Additional model details are provided in Section~S.3 of the Supplementary Material. The mean test-set PR AUC was 0.87 (SD = 0.02).


Prediction of user-level susceptibility, while accounting for the additional covariates used for matching, is obtained through an unconditional logistic regression model that regresses the susceptibility indicator on the matching covariates and includes the selected features, together with their estimated regression coefficients, as an offset. Across the 10 splits, we obtain an average AUC PR of 0.78 and SD of 0.01. For comparison, applying SoCIFR within an unconditional logistic regression framework yields mean PR AUCs (SD) of 0.78 (0.01), 0.79 (0.01), and 0.77 (0.01) under the linear (M1), additive (M2), and nonlinear (M3) specifications, respectively. The two random forest approaches based on unconditional models achieve mean PR AUCs of 0.81 (0.01) for A1 and 0.79 (0.01) for A2. The competitive predictive performance of the unconditional model suggests that the covariates used for matching have only a modest association with the outcome. This conclusion is further supported by the similarity of the regression coefficient estimates from the conditional and unconditional models, as well as the near-zero estimated effects of the matching variables (see Section S.3 of the Supplementary Material). Taken together, these results provide empirical evidence that the odds ratio relating susceptibility to the posting features is approximately collapsible with respect to the matching variables \citep{Greenland1999}.

Sensitivity analyses showed little effect of the PVE threshold. Reducing the threshold from 0.95 to 0.90 decreased the number of selected features from 11 to 10, with only a slight decline in performance for the feature-based RF model (A2; mean PR AUC = 0.78, SD = 0.01). Increasing the threshold to 0.99 yielded approximately 12 features and only a modest improvement for the nonlinear SoCIFR model (M3; mean PR AUC = 0.78, SD = 0.01).

\begin{figure}
    \centering
    \includegraphics[width=\linewidth]{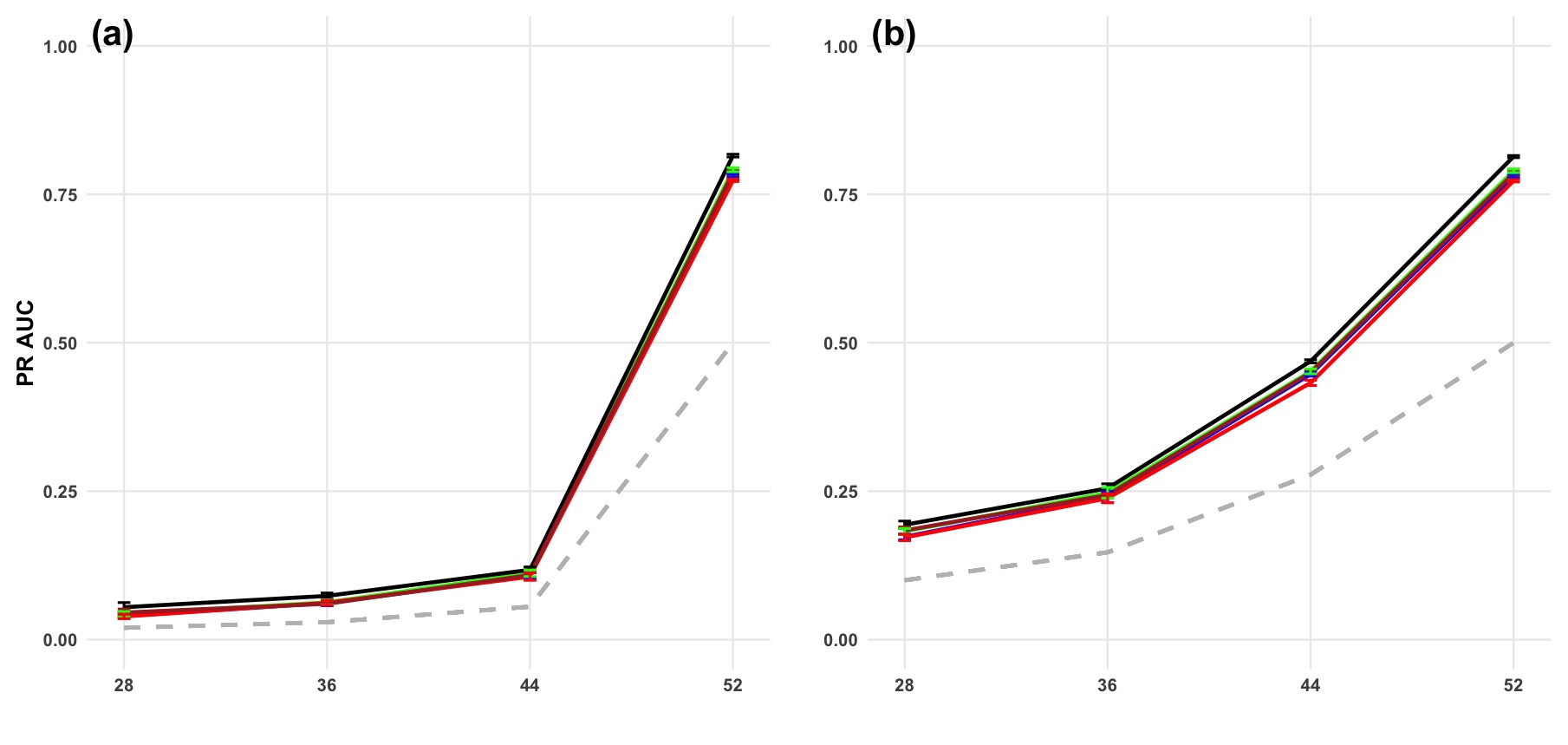}
    \includegraphics[width=\linewidth]{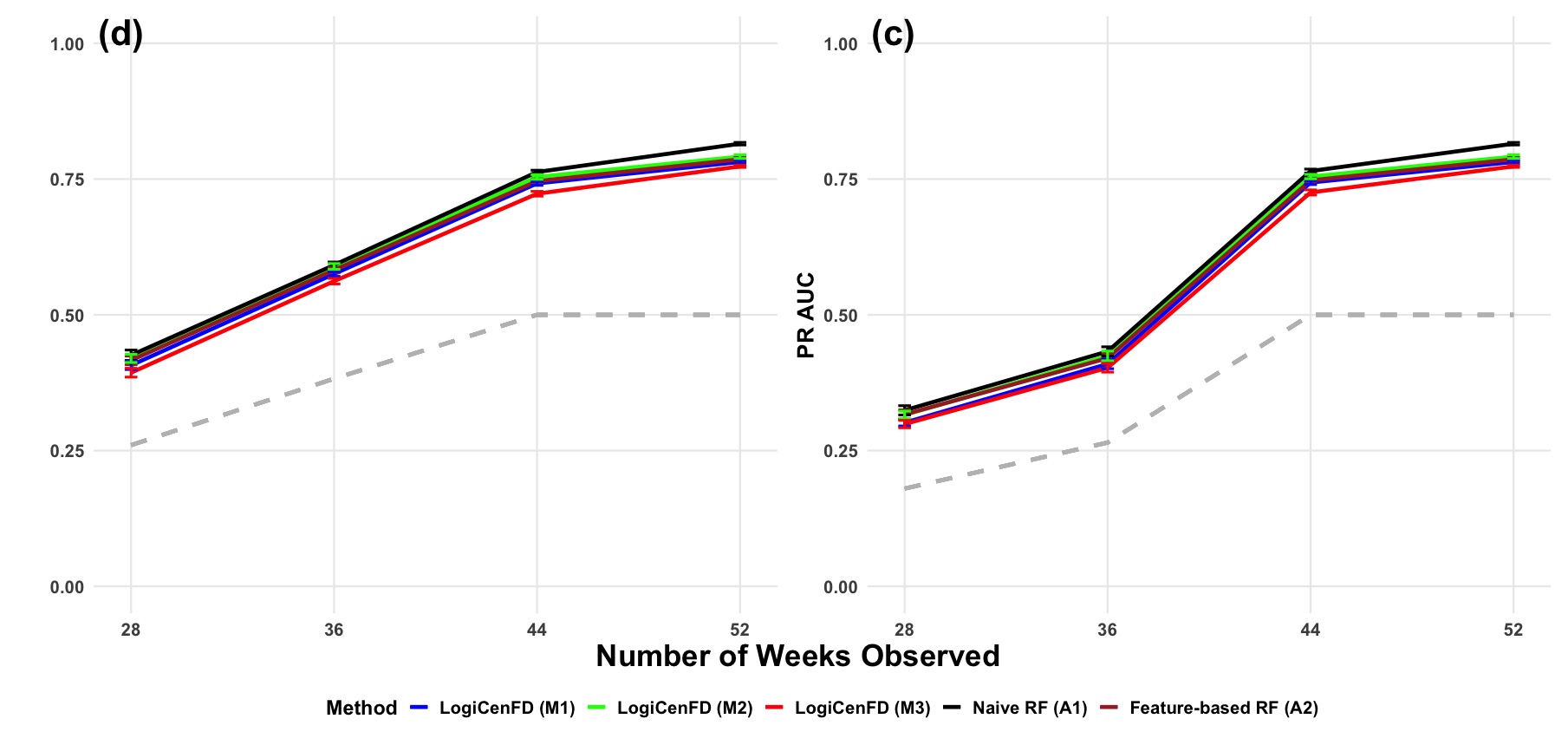}
    \caption{Mean AUC PR for the proposed SoCIFR models (M1--M3) and random forest baselines (A1, A2) across observation windows and prediction horizons. The x-axis shows the observation window length (weeks), while panels correspond to prediction horizons of \(H=0\) (a), \(H=4\) (b), \(H=8\) (c), and \(H=12\) (d) weeks following the observation period. Results are averaged over 10 random 90--10 train-test splits. The gray dashed line indicates the average prevalence of susceptible users in the test set.}
    \label{fig:prAUCPlot}
\end{figure}

To address the second goal, the dataset is redefined accordingly. Specifically, suppose the observation window consists of \(m\) weeks, selected within the available 52-week period, and the objective is to predict susceptibility to bot interaction within a future horizon of \(H\) weeks following the observation period. The outcome is then defined by whether the account interacts with a bot during this future horizon. As a result, accounts previously labeled as susceptible may or may not retain that label, depending on whether a bot interaction occurs within the next \(H\) weeks. In contrast, accounts originally labeled as controls remain controls, since they do not interact with bots for several years after the observation period. This construction has two implications: (1) the original matched case-control structure is no longer preserved, since each pair is no longer guaranteed to contain one case and one control; and (2) the prevalence of susceptible accounts is reduced. We predict susceptibility using the proposed SoCIFR framework with a logistic link under an unconditional modeling approach. Figure~\ref{fig:prAUCPlot} displays the mean out-of-sample PR AUC for the three SoCIFR variants across 10 random splits of the full matched dataset. The results correspond to observation windows of length $m$, ranging from 28 to 52 consecutive weeks randomly selected within the original one-year observation period, with susceptibility assessed over the subsequent $H$-week prediction horizon. The results indicate that the proposed SoCIFR methodology achieves predictive performance comparable to that of random forests while offering substantially greater interpretability. In particular, the estimated coefficient functions provide insight into how individual behavioral measures contribute to susceptibility to bot interaction. The results suggest that all four longitudinal features are associated with bot interaction risk. Among them, tweet volume and tweet length exhibit increasingly positive associations toward the end of the observation period, whereas posting activity and the capped ${@}$-tweet rate display more complex, non-monotonic relationships with susceptibility.

Across all settings, the proposed method performs competitively with random forests while retaining interpretability under linear and additive specifications. It accommodates both regular and irregular sampling designs and extracts low-dimensional features that capture dependence within and across covariates. These features naturally extend to matched studies through a conditional pair-based framework, an extension not directly available to a naive random forest.


\section{Discussion}\label{sec:discussion}

This paper proposes logistic functional regression models for predicting social media users’ susceptibility to interacting with automated accounts. Beyond its methodological contributions, the motivating application introduces a framework for identifying users at risk of engaging with fraudulent or automated accounts, an important departure from prior work, which has focused primarily on detecting bot accounts themselves. The proposed approach also enables prediction of susceptibility at arbitrary future time points based on posting behavior observed over windows of varying lengths, providing a flexible tool for estimating a user’s future risk of bot interaction.
From a methodological perspective, we develop a novel framework for extracting informative low-dimensional features from functional data observed under an informative sampling design and extend existing methods for censored functional data to explicitly account for this setting. In the presence of additional covariates, we propose an efficient integration strategy that is particularly attractive when multiple functional covariates share the same informative sampling mechanism. While recent work has focused on estimation of mean and covariance functions for censored functional data \citep{murphy2026}, our approach additionally accounts for the informative observation process through inverse probability weighting. By modeling the latent processes as mixtures of Gaussian processes, we obtain low-dimensional representations of user behavior via multivariate functional principal component scores, which are then incorporated into logistic regression models for predicting bot interaction susceptibility.


Future work includes developing more flexible multivariate FPCA methods for mixed-type functional data \citep{dey2026} observed under informative design and extending the framework to matched-pair settings with within-pair dependence. While the working independence assumption appears adequate here, accounting for pairwise dependence would further broaden the applicability of the methodology.

\section*{Data Availability}

The original data were obtained using the Twitter API and can be provided by the authors upon request. A cleaned version of the dataset used in the analyses is provided in the Supplementary Material.

The authors report there are no competing interests to declare. 



\section*{Generative AI and AI-assisted technologies}
AI/ChatGPT was used to check the grammar and shorten the content during revision. After using this tool/service, the authors reviewed and edited the content as needed and take full responsibility for the content of the published article.

\small
\spacingset{1.0}



\bibliographystyle{rss.bst}
\bibliography{reference}

\end{document}

%% file: commands.tex
\newcommand{\btheta}{ \mbox{\boldmath $\theta$}}
\newcommand{\bmu}{ \mbox{\boldmath $\mu$}}
\newcommand{\balpha}{ \mbox{\boldmath $\alpha$}}
\newcommand{\bbeta}{ \mbox{\boldmath $\beta$}}
\newcommand{\bmeta}{ \mbox{\boldmath $\eta$}}
\newcommand{\bdelta}{ \mbox{\boldmath $\delta$}}
\newcommand{\blambda}{ \mbox{\boldmath $\lambda$}}
\newcommand{\bgamma}{ \mbox{\boldmath $\gamma$}}
\newcommand{\brho}{ \mbox{\boldmath $\rho$}}
\newcommand{\bpsi}{ \mbox{\boldmath $\psi$}}
\newcommand{\bepsilon}{ \mbox{\boldmath $\epsilon$}}
\newcommand{\bomega}{ \mbox{\boldmath $\omega$}}
\newcommand{\bDelta}{ \mbox{\boldmath $\Delta$}}
\newcommand{\bSigma}{ \mbox{\boldmath $\Sigma$}}
\newcommand{\bOmega}{ \mbox{\boldmath $\Omega$}}
\newcommand{\bGamma}{ \mbox{\boldmath $\Gamma$}}

\newcommand{\bxi}{ \mbox{\boldmath $\xi$}}
\newcommand{\bzeta}{ \mbox{\boldmath $\zeta$}}
\newcommand{\bU}{ \mbox{\bf U}}

\newcommand{\bA}{ \mbox{\bf A}}
\newcommand{\ba}{ \mbox{\bf a}}
\newcommand{\be}{ \mbox{\bf e}}
\newcommand{\bP}{ \mbox{\bf P}}
\newcommand{\bx}{ \mbox{\bf x}}
\newcommand{\bX}{ \mbox{\bf X}}
\newcommand{\bB}{ \mbox{\bf B}}
\newcommand{\bZ}{ \mbox{\bf Z}}
\newcommand{\by}{ \mbox{\bf y}}
\newcommand{\bY}{ \mbox{\bf Y}}
\newcommand{\bz}{ \mbox{\bf z}}
\newcommand{\bh}{ \mbox{\bf h}}
\newcommand{\bmm}{ \mbox{\bf m}}
\newcommand{\br}{ \mbox{\bf r}}
\newcommand{\bt}{ \mbox{\bf t}}
\newcommand{\bs}{ \mbox{\bf s}}
\newcommand{\bb}{ \mbox{\bf b}}
\newcommand{\bff}{ \mbox{\bf f}}
\newcommand{\bL}{ \mbox{\bf L}}
\newcommand{\bu}{ \mbox{\bf u}}
\newcommand{\bv}{ \mbox{\bf v}}
\newcommand{\bV}{ \mbox{\bf V}}
\newcommand{\bG}{ \mbox{\bf G}}
\newcommand{\bC}{ \mbox{\bf C}}
\newcommand{\bg}{ \mbox{\bf g}}
\newcommand{\bH}{ \mbox{\bf H}}
\newcommand{\bI}{ \mbox{\bf I}}
\newcommand{\bD}{ \mbox{\bf D}}
\newcommand{\bM}{ \mbox{\bf M}}
\newcommand{\bW}{ \mbox{\bf W}}
\newcommand{\bw}{ \mbox{\bf w}}
\newcommand{\bT}{ \mbox{\bf T}}
\newcommand{\bQ}{ \mbox{\bf Q}}
\newcommand{\bR}{ \mbox{\bf R}}
\newcommand{\bfe}{ \mbox{\bf e}}
\newcommand{\bzero}{ \mbox{\bf 0}}

\newcommand{\iid}{\stackrel{iid}{\sim}}
\newcommand{\indep}{\stackrel{indep}{\sim}}
\newcommand{\Xtil}{{\tilde X}}
\newcommand{\bXtil}{\bf {\tilde X}}

\newcommand{\KL}{Karhunen-Lo\`eve }

\newcommand{\myT}{{\mathcal T}}

\newcommand{\calR}{{\cal R}}
\newcommand{\calL}{{\cal L}}
\newcommand{\calG}{{\cal G}}
\newcommand{\calD}{{\cal D}}
\newcommand{\calC}{{\cal C}}
\newcommand{\calS}{{\cal S}}
\newcommand{\calB}{{\cal B}}
\newcommand{\calA}{{\cal A}}
\newcommand{\calT}{{\cal T}}
\newcommand{\calO}{{\cal O}}
\newcommand{\calN}{{\cal N}}
\newcommand{\calK}{{\cal K}}
\newcommand{\calH}{{\cal H}}
\newcommand{\calX}{{\cal X}}

\newcommand{\argmax}{{\mathop{\rm arg\, max}}}
\newcommand{\argmin}{{\mathop{\rm arg\, min}}}
\newcommand{\Frechet}{\mbox{Fr$\acute{\mbox{e}}$chet}}
\newcommand{\Matern}{ \mbox{Mat$\acute{\mbox{e}}$rn}}
\newcommand{\Prob}{\mbox{Prob}}

\newcommand{\beq}{\begin{equation}}
\newcommand{\eeq}{\end{equation}}
\newcommand{\beqn}{\begin{eqnarray}}
\newcommand{\eeqn}{\end{eqnarray}}

\newcommand{\beqnno}{\begin{eqnarray*}}
\newcommand{\eeqnno}{\end{eqnarray*}}